\documentclass[fleqn,10pt]{wlscirep}
\usepackage{color}
\usepackage{bm}
\usepackage{float}

\title{Surface temperatures in New York City: Geospatial data enables the accurate prediction of radiative heat transfer}

\author[1,2]{Masoud Ghandehari}
\author[3,4,*]{Thorsten Emig}
\author[1]{Milad Aghamohamadnia}
\affil[1]{New York University, Tandon School of Engineering, Brooklyn,
  NY, USA}
  \affil[2]{New York University, Center for Urban Science and Progress, Brooklyn, NY, USA}
\affil[3]{Massachusetts Institute of Technology, MultiScale Materials Science
for Energy and Environment, Joint MIT-CNRS Laboratory UMI 3466,
Cambridge, Massachusetts 02139, USA}
\affil[4]{Laboratoire de Physique
Th\'eorique et Mod\`eles Statistiques, CNRS UMR 8626, B\^at.~100,
Universit\'e Paris-Sud, 91405 Orsay cedex, France}

\affil[*]{emig@mit.edu}

\begin{abstract}
Three decades into the research seeking to derive the urban energy budget, the dynamics of the thermal exchange between the densely built infrastructure and the environment are still not well understood. We present a novel hybrid experimental-numerical approach for the analysis of the radiative heat transfer in New York City. The aim of this work is to contribute to the calculation of the urban energy budget, in particular the stored energy. Improved understanding of urban thermodynamics incorporating the interaction of the various bodies will have implications on energy conservation at the building scale, as well as human health and comfort at the urban scale. The platform presented is based on longwave hyperspectral imaging of nearly 100 blocks of Manhattan, and a geospatial radiosity model that describes the collective radiative heat exchange between multiple buildings. The close comparison of temperature values derived from measurements and the computed surface temperatures (including streets and roads) implies that this geospatial, thermodynamic numerical model applied to urban structures, is promising for accurate and high resolution analysis of urban surface temperatures. 
\end{abstract}

\begin{document}

\flushbottom
\maketitle

\thispagestyle{empty}


\section*{Introduction}

Cities are home to the majority of the world’s population and thus significantly determine global energy consumption, waste, and pollution. The dynamics of the urban energy budget, especially the thermal exchange between the densely built infrastructure and the surrounding environment, are not well understood. This is largely because the component of the energy budget associated with energy storage has been unattainable.  The significant of this gap became was highlighted in the early 1990's through classic contributions by number of researchers working on the derivation of the urban energy budget; which included work on the urban heat island, radiative heat transfer and the stored energy. \cite{23,Oke:1987mo,Oke:1988mf,Grimmond:1991yb,Roth:1994hy}. This body of work was subsequently expanded to incorporate urban scale climate models that included the application of  satellite remote sensing, resulting in better understanding of the thermal dynamic responses of the urban environments \cite{Salamanca:2010tt,Heiple:2008kr,Martilli:2003hc,Chen:2011xq,Kato:2007ad,15,Roberts:2006lk,Offerle:2005sh,Kusaka:2001mj}. Nonetheless, the quantitative analysis of the thermal storage component is still elusive. This is because of the large number of unknowns in the urban space equations of heat transfer. Time resolved analysis of the urban surface temperature is perhaps the most effective avenue for closing this knowledge gap. Advancing the understanding of the energy budget will lead to improvements in several areas: models of urban meteorology and air quality, models that forecast energy demand and consumption, technological innovations in building materials, heating and cooling technologies, as well as climate control systems and urban design, all of which seek to enable energy efficiency at the building level and at city scale, while improving human health and the quality of the environment. 
When considering the state of art approaches to measuring urban surface temperatures, there has been number of challenges:

\begin{itemize}
\item Temperatures of vertical surfaces, which constitute the main portion of urban building surfaces, are inaccessible by satellite and/or aerial remote sensing. This information is essential in the accurate prediction of the urban storage flux.
\item In the case of ground based remote sensing, the unknown surface emissivity plays a crucial role in the accurate determination of surface temperatures. Comprehensive databases of surface emissivity values for buildings in cities are not readily available.
\item The radiation measured by longwave sensors is combination of gray body radiation of the target, as well as the radiation reflected (diffuse and specular) from other surfaces. Therefore measured values do not always  represent the intrinsic radiation of the target. Separation of the reflected portion is required for accurate assessment of the surface temperature.
\item The atmospheric constituents have significant effect in the interpretation of the measured values and the accuracy of numerical models when using actual values of sky radiance. There are number of approaches to compensate for the atmospheric effects.
\end{itemize}

The research presented in this article seeks to address some of these challenges. Here, we present results of studies done in New York City mapping the surface radiations from nearly 100 blocks of Manhattans West Side. This includes measurements using a hyperspectral imaging (HSI) instrument, and a numerical model for calculating the measured radiation. The model results are subsequently compared with the measured values. This work benefits from a legacy of applications of spectroscopic imaging in earth sciences and remote sensing, including surface radiography and plume detection \cite{Pearlman:2013iz,Tratt:2014aa}.  In the majority of those applications, imaging systems are deployed in a “downward-looking” configuration, mounted on moving platforms such as aircraft and satellites. In contrast, when considering urban energy research, stationary ground based imaging offers the advantage of persistence and a desirable field of view. Oblique view urban imaging has shown promise, for example for mapping the persistent leakage of refrigerant gases from a large number of structures in New York City \cite{Ghandehari:2017aa}.

\section*{Results}

Surface radiance Measurements were carried out using a long wave hyperspectral instrument. Buildings were imaged along an eight kilometer portion of NYC's West Side (Figures 1 and 2). This was followed by a numerical simulation of long wave radiosity of the scene (Figure 3). Measurements were carried out by installing the instrument on a rooftop vantage point at Stevens Institute of Technology in Hoboken, New Jersey. This provided an unobstructed view of Manhattan’s West Side with views including both low and high rise structures. Images were collected at 128 spectral bands, and for one week at cadences ranging from 10 seconds to 3 minutes. The 1.1 mrad angular resolution, when applied to target distances from 1 to 5 km, resulted in the spatial resolution ranging from 1.1-5.5 meters per pixel. This spatial scale can enable the incorporation of useful attributes at building level (including fuel type, composition, occupancy, etc.), not used at this stage, but an option when the platform is at higher level of maturity. 
The imaging system was manufactured by the Aerospace Corporation. The focal plane has sensors with spectral response ranging from 7.6–13.2 $\mu$m; this corresponds to peaks of blackbody spectra at temperatures between 230K and 395 K, thereby useful for the estimation of urban solid surface temperatures. This spectral response also includes a region in which many polyatomic molecules (including gaseous emissions) have well-defined spectral features. Those features including the 40 nm spectral resolution and the Noise Equivalent Spectral Radiance (NESR) of ~1 $\mu$Flick ($\mu$W cm$^{-2}$ sr$^{-1}$ $\mu$m$^{-1}$) also allow gaseous compounds to be identified with high selectivity \cite{Ghandehari:2017aa}.

\subsection*{Derivation of Surface Temperature and Emissivity from Hyperspectral Imagery}

\begin{figure}[th]
\centering
\includegraphics[width=\linewidth]{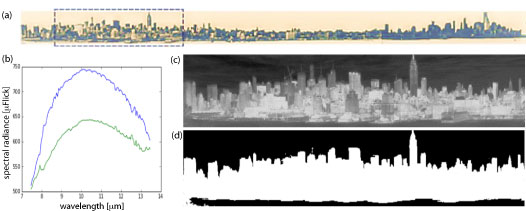}
\caption{(a) Scan of eight kilometers along Manhattan West side 
(image by python https://codeshare.io/5ONkBP script, one wavelength from data cube)  (b) Bi-cluster center spectrum. The blue curve represents
  building surfaces and the green curve represents sky and
  water. Wavelength in micrometers (horizontal) and radiance in micro
  flicks (vertical); (c) Brightness temperature image, averaged over
  all bands of hyperspectral data cube; (d) masked area of buildings.}
\label{fig:Bi-cluster}
\end{figure}

The spectral emission from a body is a function of wavelength, the body's surface emissivity and its surface temperature. While the emissivity depends on the wavelength, the temperature is unique for an object at a given instance of time. A challenge in the application of thermographic imaging for mapping surface temperatures in heterogeneous terrains (such as buildings in a city) is that the surface emissivity is often unknown. That emissivity may range from ~0.3 (aluminum and “low-e” glass) to higher values of 0.9 (concrete, brick, and polymeric composites). In wavelengths between 7-14 micron, a 1$\%$ change in emissivity will result in approximately 0.5$^\circ$ K temperature difference. Therefore, the possible full range of emissivity values for building façades can correspond to as high as 30$^\circ$ K temperature. An additional challenge is that the measured radiance at each pixel also includes effects of atmospheric scattering, absorption or emission by ambient gases along the path of observation. Therefore, when considering applications of thermography for studying urban thermodynamics, e.g. building energy and the urban heat island, significant errors may result if the emissivity effect is ignored. 
Hyperspectral imagery has been applied for the separation of the effect of emissivity versus surface temperature when considering measured radiance. These approaches have largely been applied to downward looking (aerial and satellite) remote sensing. Some techniques use calibrated curves of emissivity derived from laboratory experiments, in combination of instrumented measures of atmospheric parameters (e.g. using ASTER satellite, Gillespe. Pivovarnik); others achieve the separation as a byproduct of the process for calculation of the atmospheric effects. Overall, calcualtion for the compensation of effect of the atmosphere is done following two distinct approaches: 1) using measured values of atmospheric constituents followed by modeling, the most commonly being the Moderate Resolution Atmospheric Transmission (MODTRAN) \cite{Berk:aa}, and 2) using radiance values measured by the hyperspectral sensor, without the use of auxiliary data, a technique commonly known as "In Scene Atmospheric Compensation" (ISAC) \cite{Young:2002ea}. In this study we use the ISAC algorithm proposed by Young et. al \cite{Young:2002ea} for the calculation of the atmospheric effects, which also includes the separation of temperature versus emissivity. We also compared the ISAC approach with results obtained from the MODTRAN model which incorporates measured ambient concentrations of gases. Some background and brief summary of the approach and results are given below. 
In preparation for the application of ISAC algorithm following preprocessing was done \cite{Gillespie:1998rr,Pivovarnik:2017fu}:

\begin{enumerate}
\item Conversion of raw telemetry data to engineering data
\item Radiometric and geometric calibrations
\item Bad pixel replacements
\item Spectral Smile Removal
\end{enumerate}

\begin{figure}[ht]
\centering
\includegraphics[width=\linewidth]{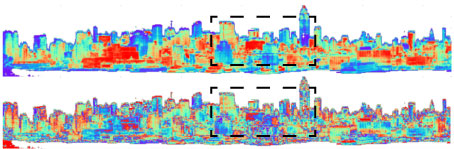}
\caption{(Top) Temperature, and (bottom) emissivity map of the scene
  10$\mu$m wavelength. The framed region marks the area for which
  surface temperatures have been computed by the radiosity method.}
\label{fig: T+emissivity_maps}
\end{figure}

In contrast with the downward looking imaging, application of ISAC to the oblique view ground based imaging  should be considered with two caveats. 1) The first is the significance of the unequal distance from the sensor to the object, and the corresponding effect on the automated atmospheric compensation, which is inherent in the ISAC technique. In applications involving aerial or satellite platforms the object to sensor distance does not vary greatly, however in the oblique configuration, the distance can vary significantly. In our case (considering the scene in Figure \ref{fig:Bi-cluster}) it is from 1 to 2 km. On the positive side, the smaller working distance in ground based sensing corresponds to a relatively smaller atmospheric effect. 2) The second point that needs to be considered is that ISAC relies on approximately 10$\%$ of the scene to be occupied by target emissivity close to 1. In downward looking remote sensing, these are often bodies of water which satisfy this condition. However, in the oblique view imaging, bodies of water (Hudson river in our case) are not ideal because of the ripple effect observed in the oblique angle diminishes the high emissivity advantage of water when applied to the ISAC algorithm. As a compromise, the large portion of high emissivity pixels corresponding to building materials composed of calcium (e.g. limestone), Silica (e.g. Bricks), or calcium silicates (e.g. concrete), with emissivity as high as 0.95, will to a great extent serve the black body requirement of the Algorithm. The error caused by this assumption was indirectly evaluated by spot checking the surface temperature calculations at selected pixels, using the MODTRAN model \cite{Berk:aa}, incorporating local measurement of atmospheric constituents \cite{Berk:aa}. 
The calculations were done by first masking the building region of the scene (Figure \ref{fig:Bi-cluster}) for the application of ISAC algorithms \cite{Young:2002ea}, followed by the Temperature Emissivity Separation (Figure \ref{fig: T+emissivity_maps}) . Masking was done using K-Means clustering with two clusters. One was the sky, clouds and water pixels as a cluster with lower temperature, i.e., lower radiance. The other are the buildings as a cluster being hotter thus higher radiance. The two cluster mean spectrum associated with the brightness temperature of the scene and the masked image are shown in Figure \ref{fig:Bi-cluster}.

Detailed calculation of atmospheric effects at selected pixel groups was carried out using the MODTRAN model \cite{Berk:aa}.  The model incorporates the radiometric effects from the sky (e.g., reflection) as well as the prevailing atmospheric effects in the oblique view of the scene (Meler 2011).  The model incorporated twenty-eight input parameters.  Twenty-five of these were the atmospheric concentrations of various gases, twenty-two of which were taken from conditions typical in a mid-latitude summer air column (see Table \ref{tab:correction}), while (H$_2$O, O$_3$, CO$_2$) concentrations were measured ambient values obtained from local weather stations (Table \ref{tab:compensation}).  The remaining three parameters were air temperature, background surface temperature, and emissivity.  Air temperature was obtained from local weather stations records. This approach follows the atmospheric correction procedures used for airborne HSI instruments such as ATLAS or HysPIRI, where radiosound launches are used to determine the gas concentrations. In our case the calculation of the background surface temperature requires the variable path length from sensor to the target location on the building surface. This was obtained using a 3D digital surface model (DSM) of the city derived from the NYC building data \cite{NYCPlanning2016}.
Other than the above atmospheric effects, the radiance recorded by the sensor also includes the sky radiance and scattered radiation along the observation path (known as down-welling radiation in airborne platforms). The calculation of this downwelling radiation is challenging in the case of oblique view remote sensing. Nonetheless, certain assumptions can simplify the calculations. One is the use of similarity between the night-time radiation received by the objects in the scene, versus radiation received by the sensor from the sky. The building surfaces in the scene are vertically aligned and receive the sky radiance over a range of 90 degrees (from vertical to horizontal). The sensor swath width and height are 94 and 6 degrees, respectively. The field of view looking from the instrument, from Hoboken NJ, toward Manhattan is occupied by approximately 50$\%$ sky, 40$\%$ buildings and 10$\%$ river. The same can be said if the sensor was to be located in Manhattan looking toward Hoboken. Therefore, it can be assumed that the average of all pixels received in the 6 $\times$ 94 degree window is representative of what an object receives from its surroundings and seen by the sensor.  This "in-scene" measurement was used to approximate the downwelling radiance. The resulting values of surface temperature derived by MODTRAN was subsequently compared the ISAC results for selected pixels in a high rise area and a low rise part of the city (Figure 3). Results of the comparison are quite good for this preliminary stage of work (Table 3). This comparison will need to be automated and applied to the entire scene in order to arrive at the statistics of the difference between the two approaches.

\begin{table}[ht]
\centering
\begin{tabular}{|l|c|}
\hline
Compound & Concentration [g/cm$^2$.m]  \\
\hline
F11 & 2.93E-11 \\
F12 &5.03E-11 \\
CCL3F &2.10E-19 \\
CF4 & 2.10E-19 \\
F22 &1.26E-11 \\
F113 &3.98E-12 \\
F114 &2.51E-12 \\
R115 &2.10E-19 \\
CLONO2 &1.21E-12 \\  
HNO4 & 9.10E-14 \\
CHCL2F & 2.10E-19 \\
CCL4 & 2.72E-11 \\
N2O5 & 5.08E-17 \\
CO & 3.14E-08 \\
CH4 & 3.56E-07 \\
N2O & 6.71E-08 \\
O2 & 4.38E-02 \\
NH3 & 1.05E-10 \\
NO & 6.29E-11 \\
NO2 & 4.82E-12 \\
SO2 & 6.29E-11 \\
HNO3 & 1.05E-11 \\
\hline
\end{tabular}
\caption{\label{tab:correction} Concentration of trace gases used for modeling the atmospheric effect.}
\end{table}

\begin{table}[ht]
\centering
\begin{tabular}{|l|l|}
  \hline
  Timestamp & 04/13 15:22 EST \\
  \hline
Water vapor & 1.47 g/cm$^3$ \\
Ozone &0.00026 g/cm$^3$ \\
Carbon Dioxide & 400 ppmv \\
\hline
\end{tabular}
\caption{\label{tab:compensation} Measured values used for modeling the atmospheric effect.}
\end{table}

\begin{table}[ht]
\centering
\begin{tabular}{|l|l|l|}
  \hline
  & (ISAC) & (MODTRAN) \\
  \hline
Low Rise Bldg     &   286.3 & 285.7 \\
High Rise Bldg     &  284.1 & 283.3 \\
\hline
\end{tabular}
\caption{\label{tab:surface_T} Comparison of surface temperature derived using ISAC versus MODTRAN (in Kelvin).}
\end{table}

\subsection*{Comparison to Radiosity Model Predictions}

in order to enable studies on the radiative heat transfer between urban structures (here buildings and streets) we have employed the radiosity method to compute the heat radiation emitted and reflected from all surfaces, including multiple reflections (see Methods). Our model for the urban geometry is derived by surface triangulation from a geospatial dataset (see Methods). For each triangle of the surface mesh, we define its emissivity $\epsilon$, wall thickness $d$, thermal conductivity $\kappa$, and the temperature $T^\text{int}$ on the inside of the surface wall. This information together with the long wavelength radiant flux $L$ from the sky determines uniquely the outside surface temperatures in equilibrium for all triangular surface elements. Equilibrium refers to the assumption that there is no incoming radiant flux that varies over time, like day time solar radiation. Hence, we expect that our model can predict surface temperatures in the evening and during early morning before sunrise, or in strong cloud covered conditions.  
Without prior knowledge of the building envelopes composition and interior temperatures, we have assumed typical value of $\epsilon_\text{wall, roof}=0.95$, $\kappa_\text{wall, roof}=1.05$W/(K m), $T^\text{int}_\text{wall, roof}=293.15^\circ $K for walls and roofs, and $\epsilon_\text{street}=0.93$, $\kappa_\text{street}=1.25$W/(K m), $T^\text{int}_\text{street}=283.15^\circ $Km for streets and $d=0.2$m for all surfaces.  Long wave length radiant flux from the sky was estimated to be $L=300$W/m$^2$. These estimates are made as a baseline to establish a platform by which the radiant flux of the city can be analyzed at both the building level and the city level, simultaneously. Additional information on the building envelop materials would naturally improve the results. It is expected that with increase of the availability of urban data, building information will also become more available in the near future, contributing to the model acuity.

\begin{figure}[th]
\centering
\includegraphics[width=\linewidth]{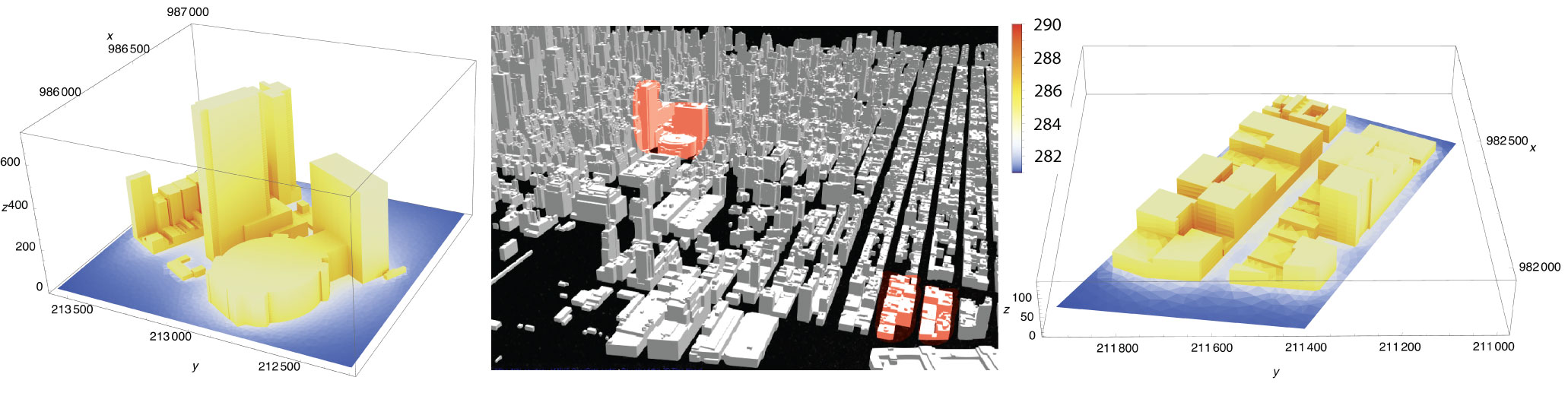}
\caption{3D map of the area including modeled blocks of buildings (marked in red: 3 high rise
  blocks, upper left and 2 low rise, lower right) studied for model
  verification, and 3D visualization of computed surface temperatures
  (see text for details). Linear dimensions on the frame axes are in
  feet, temperatures in Kelvin.}
\label{fig:NYC_map}
\end{figure}

\begin{figure}[th]
\centering
\includegraphics[width=0.7\linewidth]{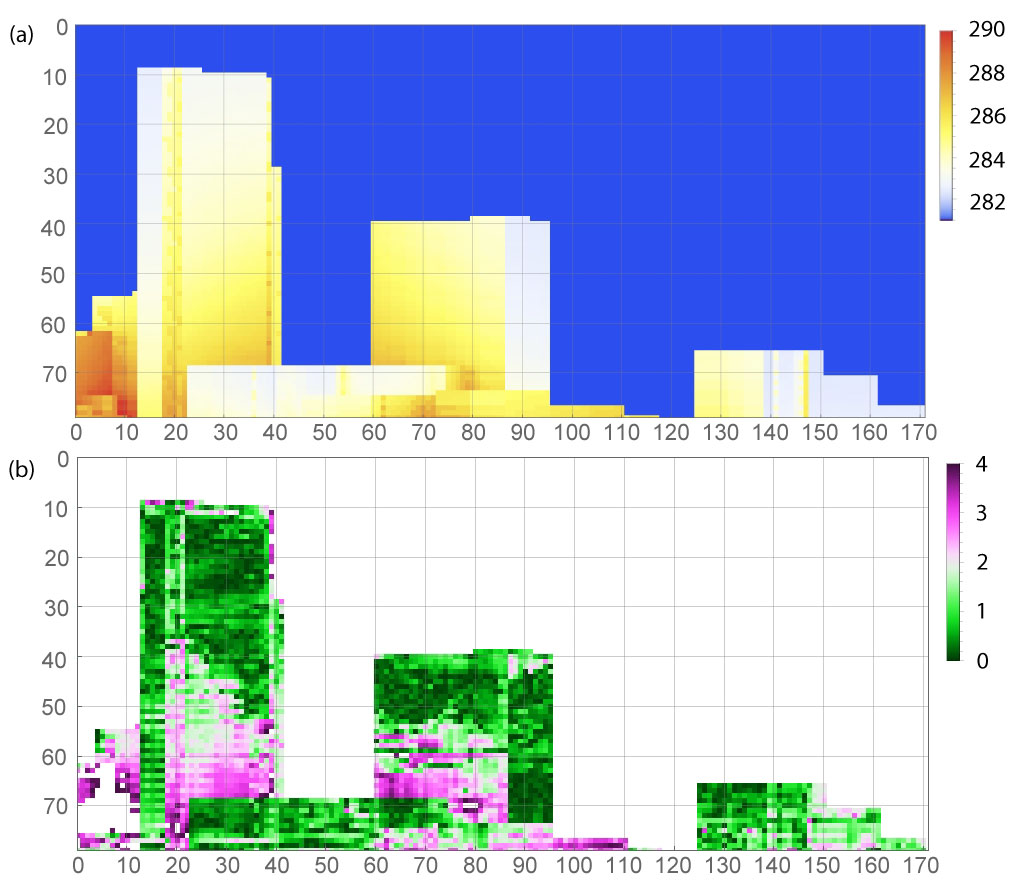}
\caption{(a) Two-dimensional projection of computed surface temperatures for the two regions shown in Figure \ref{fig:NYC_map}. (c) Temperature
  difference between (a) and experimentally determined surface temperatures, only for building surfaces.  All
  temperatures are in Kelvin; the spatial dimensions are measured in camera
  pixels.}
\label{fig:T_map_exp}
\end{figure}

We compared the measured and model results for two block groups in Manhattan (Figure 3), one is largely composed of high rise buildings (area HR) and the other composed of low rise buildings (area LR). Area HR consists of three blocks bounded by 7th to 8th Avenue~and W 31st to 35rd  Street, and area LR consists of two blocks bounded by 10th to 11th Avenue~and W 20th to 22nd Street. Figure \ref{fig:NYC_map} shows the isometric bird eye view of the blocks, and Figure 4 is the virtual view of the same two block groups from the instrument location. The linear dimensions (in feet) of the two areas is indicated on the 3D views. A total number of 31339 triangular surface elements have been used to model the two areas. A 3D view of the resulting surface temperatures in shown in Figure \ref{fig:NYC_map} for the two areas, with color coded temperature, and the same scale as in Figure \ref{fig:T_map_exp}(a).

The 3D computed surface temperatures for the high rise and low rise areas is projected on the two-dimensional plane using the same view angle as the hyperspectral imager. it also has a pixel resolution that corresponds to the imager (1.1 rad per pixel in horizontal and vertical directions). Figure \ref{fig:T_map_exp} shows the pixel matrix of the computed temperatures. Since the model describes only urban surfaces, the sky temperature is set to the average measured value. The HR and LR areas contain a large fraction of the building surfaces visible from the observation position. There are also surfaces of buildings between the two modeled areas that partially mask the buildings visible in Figure \ref{fig:T_map_exp}(a). Computed and measured surface temperatures can be compared by studying the absolute value of the difference between the that derived from measurements and the modeled values of temperature, For the selected scene pixels visible by the imager, the agreement between the measured and computed values is very satisfying as shown in Figure \ref{fig:T_map_exp}(b), with the lower bound of the difference at  ~1$^\circ$K (for the majority of pixels) to larger deviations of ~3$^\circ$K for the lower floors which has greater number of multi path reflections.

\begin{figure}[ht]
\centering
\includegraphics[width=0.6\linewidth]{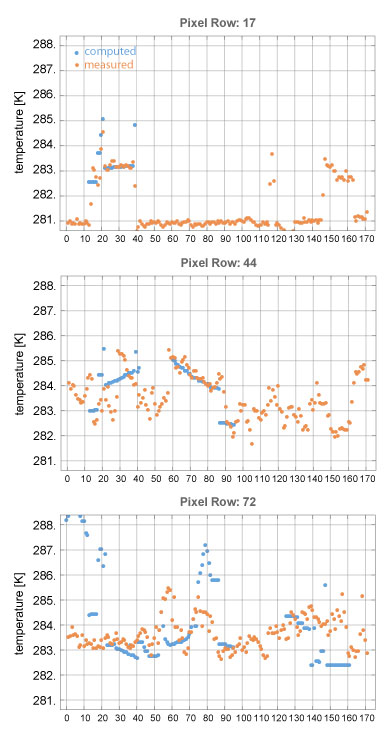}
\caption{Measured and computed surface temperatures at three different
elevation levels (pixel rows in Figure \ref{fig:T_map_exp}), as function
of camera pixels. Note that the measured data include also sky
temperatures in the gap between buildings.}
\label{fig:T_Temp_Rows}
\end{figure}

A more detailed comparison between computed temperatures and that derived from measurement is shown in Figure \ref{fig:T_Temp_Rows}. Three vertical elevations (camera pixel rows, cf.~Figure \ref{fig:T_map_exp}) are considered for comparison. Panel (a) corresponds to an elevation that cuts only the tallest building (between pixel columns 14 and 39). Measured and computed temperatures agree nicely, reproducing also the locally higher temperatures at one edge of the building, plausibly due to multiple reflections between facing walls (around pixel column 20).  Panel (b) displays a cut through the second tallest building (between pixel columns 60 and 90). This building receives heat radiation from the tallest adjacent high rise building, producing the temperature gradient across its surface. Both measured absolute temperature values and the slope of the gradient are nicely reproduced by the computed temperatures. Finally, panel (c) displays a cut at a near ground elevation. While the temperatures show substantial spatial variations due to the combined effect of many low rise buildings, the overall range of the temperature profile shows reasonable agreement between measured and computed data. The discrepancy for the high rise between pixel columns 14 and 39 is caused by a masking building. We conclude that when compared to values derived from measurement, the model captures the main features of the radiative heat transfer in complex urban geometries, even when exact material parameters like emissivity and thermal conductivity are unknown, but replaced by typical values for urban materials.

\section*{Discussion}

At the introduction we highlighted the rational for mapping the surface temperature in cities as an avenue by which the calculation of urban energy budget can be completed. The platform presented is a start in that direction, with number of additional and important steps to consider: 
\begin{enumerate}
\item Incorporation of the emissivity values derived from the measured radiance, or possibly values of emissivity from urban databases if available, can reduce errors caused by the assumptions made for the emissivity in the radiosity model.
\item In the case of using remote sensing for derivation of the surface emissivity, higher resolution imagery will significantly improve the results. This can be achieved either by smaller instantaneous field of view (IFOV), or shorter distance to the target. The current spatial resolution of 1 to 3 meters introduces errors in certain parts of the scene, when mixed pixel values are present (i.e. when a pixel represents more than one material, like windows vs. facade). it should be added that we did carry out a preliminary sensitivity test to study the effect emissivity values 0.85, 0.9, and 0.95 . The difference between the modeled and measured values (similar to that shown in Figure 4b) remained below 3 Kelvins for all the three emissivity values. There was however some differences in the distribution of temperature across the building facade.
\item The final step is the separation of the reflected from the intrinsic radiation of the target, now possible using the presented radiosity model.A high resolution radiative transfer model of a city can subsequently be obtained, which when processed in a time resolved mode, would lead to more accurate derivation of the stored energy. 
\end{enumerate}

\section*{Methods}

\subsection*{Geospatial Dataset}

Footprints of buildings in the form of two-dimensional polygons and the maximum building roof height above ground elevation were obtained from the shapefile (Name: ``building\_1015'') provided by the City of New York, Department of Information Technology and Telecommunications (https://data.cityofnewyork.us/Housing-Development/Building-Footprints/nqwf-w8eh). From these data polygons, the envelopes of connected buildings were constructed by removing internal walls between the buildings. The wall polygons obtained were divided along the vertical direction into smaller rectangles and finally triangulated together with the polygons for the roof street surfaces to obtain a mesh of triangular surface elements. The radiosity method for computing radiative heat transfers has been applied to this mesh.

\subsection*{Model for Radiative Heat Transfer and Surface Temperatures}

The model has been developed and employed previously for patterned surfaces \cite{Emig:2017aa}. Urban geometry (building and street surfaces) is represented by a mesh of small surface ``patches'' given by $N$ mutually joining triangles $P_j$, $j=1,\ldots, N$, defined over a planar base plane ($xy$-plane). The triangles are oriented so that their surface normals
${\bf n}_j$ are pointing to the outside of the buildings or towards the sky or streets. All surface normals are either normal or parallel to the base plane. Each triangle is further characterized by an emissivity $\epsilon_j$, surface thickness $d_j$, and thermal conductivity $\kappa_j$ described in the previous section. On the inside of the surface a local equilibrium interior temperature $T^\text{int}_j$ is imposed for each triangle.  We assume that the surface receives a homogeneous radiant flux $L$ from the sky. The equilibrium temperatures $T_j$ on the outside surfaces of the triangles are determined by equating the internal and external net flux densities for each triangle. The internal net flux is obtained from the stationary heat conduction equation
$q^\text{int}_j=-\kappa \partial_n T_j$ integrated across the surface thickness $d_j$ yielding $q^\text{int}_j=(T_j-T^\text{int}_j)\kappa_j/d_j$. The external net
flux $q^\text{ext}_j$ is obtained as the sum of the incoming fluxes from the sky ($L$) and those scattered from all other visible triangles and the heat flux $\sigma \epsilon_j T_j^4$ radiated by the surface $j$ where $\sigma$ is the Stefan-Boltzmann constant. 

For the case of a single planar surface ($j=N=1$), the
condition $q^\text{ext}_1=q^\text{int}_1$ yields
\begin{equation}
  \label{eq:1}
  (T_\text{flat}-T^\text{int})\frac{\kappa}{d} = \epsilon (L-\sigma
  T_\text{flat}^4) \, ,
\end{equation}
which determines the outside surface temperature $T_\text{flat}$ of the flat surface as function of known parameters. 

For an urban geometry one has to consider multiple reflections between surface patches that contribute to the net external fluxes. To describe this effect, it is assumed that the surface patches are gray diffusive emitters, i.e., the emissivity is frequency independent and the radiation density is constant across the surface patches and emitted independent of direction. This is justified since the thermal wavelengths (microns) are small compared to urban structures and hence the size of the surface patches. We apply the radiosity method \cite{Modest:ij} to obtain the external fluxes $q^\text{ext}_j$.  For a given surface patch $P_j$, the outgoing radiant flux is given by the sum of emitted thermal radiation and the reflected incoming radiation,
\begin{equation}
  \label{eq:2}
  J_j = \sigma \epsilon_j T_j^4 + (1-\epsilon_j) E_j 
\end{equation}
where we used that the reflectivity equals $1-\alpha_j$ for an opaque surface where $\alpha_j=\epsilon_j$ is the absorptivity.  How much energy two surface patches exchange via radiative heat transfer depends on their size, distance and relative orientation which are encoded in the so called view factor $F_{ij}$ between patches $i$ and $j$. $F_{ij}$ is a purely geometric quantity and does not depend on the wavelength due to the assumption of diffusive surfaces above.  It is defined by the surface integrals
\begin{equation}
  \label{eq:3}
  F_{ij} = \int_{A_i} \int_{A_j} \frac{\cos \theta_i \cos
    \theta_j}{\pi A_i |\bm{ r}_{ij}|^2} dA_i dA_j
\end{equation}
where $\theta_{i}$ is the angle between the surface patch's normal vector $\bm{ n}_{i}$ and the distance vector $\bm{ r}_{ij}$ which connects a point on patch $i$ to a point on patch $j$, and $A_{i}$ is the surface area of patch $i$ . The view factor matrix obeys the important reciprocity relation $A_j F_{ji}=A_i F_{ij}$ and additivity rule $\sum_j F_{ij} = 1$. With this geometric quantity, the radiative flux received by surface patch $j$ from all other surface patches can be expressed as $E_j = \sum_{i} F_{ji} J_i$, and one can solve Eq.~(\ref{eq:2}) for
the vector of outgoing fluxes, yielding
\begin{equation}
  \label{eq:4}
  \bm{J} =  \left[ \bm{1} - ( \bm{1} - \bm{\epsilon})
      \bm{F}\right]^{-1} \bm{J}_0 \, ,
\end{equation}
where we combined the fluxes $J_j$ from all patches into a vector $\bm{J}$ and the radiation $\sigma \epsilon_j T_j^4$ into a vector ${\bf J}_0$ to
use a matrix notation. Here $\bm{1}$ is the identity matrix and $\bm{\epsilon}$ is the diagonal matrix with elements $\epsilon_j$. To obtain the surface temperatures $T_j$ we need to compute the net heat transfer to surface patch $j$ which is given by the incident radiation $E_j$ minus the outgoing flux $J_j$, leading to the net flux $q_j^\text{ext} = \sum_i F_{ji}
J_i - J_j$. In vector notation this net flux becomes
\begin{equation}
  \label{eq:5}
  \bm{q}^\text{ext} = ( \bm{F} - \bm{1} ) \left[ \bm{1} - ( \bm{1} - \bm{\epsilon})
      \bm{F}\right]^{-1} \bm{J}_0 \, .
\end{equation}
In the stationary state, the surface patch temperatures are then determined by the condition that the net external flux equals the net internal flux, $\bm{q}^\text{ext} = \bm{q}^\text{int}$ where $\bm{q}^\text{int}$ defines the vector with elements $(T_j-T_j^\text{int})\kappa_j/d_j$ due to heat conduction across the surface. This condition uniquely fixes the temperatures $T_j$ when all other parameters including the flux $L$ from the sky are known. In the following, technically, we include the sky as an additional surface so that we have now $N+1$ surface patches. The corresponding additional matrix elements for the view factor matrix $\bm{F}$ follow from reciprocity and additivity rules, and we include the downward radiation $L$ as the $(N+1)^\text{th}$ component in $\bm{J}_0$.

\subsection*{View Factors}

The view factors $F_{ij}$ between each pair of triangles of the surface mesh are computed using a modified version of the C program View3D which is freely available under GNU General Public License at https://github.com/jasondegraw/View3D.

\subsection*{Solving the radiosity equations}

The numerical solution of the equations of the radiosity method follows these steps.

\begin{enumerate}
\item The triangular surface elements are grouped into three different classes: horizontal street patches (s), horizontal roof patches (r)   and wall patches (w) that are perpendicular the base plane and connect the patches in class s and r.

\item Computing the view factors $F_{ij}$. This needs to be done only for all patch class combinations $(w, s)$, $(w,r)$ and $(w,w)$ with   the restriction $i<j$ for $(w,w)$ since the view factors for $i>j$  follow from reciprocity. The patches of classes $s$ and $r$ cannot  see each other so that the view factor submatrix for these classes vanishes.

\item Constructing the total view factor matrix $\bm{F}$ for all triangles of classes $w$, $s$ and $r$ and the single enclosing surface describing the sky. This is done by using reciprocity to obtain  the matrix elements for the patch class combinations $(s,w)$ and   $(r,w)$. To obtain the view factor for the transfer from a surface patch $i$ towards the sky we use the sum rule $\sum_j F_{ij}=1$, i.e., $F_{i
    \,\text{sky}}= 1- \sum_{j \in \{s,r,w\}} F_{ij}$. The view factor for the transfer from the sky to a patch $i$ follows from reciprocity as $F_{\text{sky}\, i}= \frac{A_i}{A} F_{i
    \,\text{sky}}$ where $A$ is the total planar area of the urban area under consideration. 

\item The inverse matrix of Eq.~(\ref{eq:5}) can be computed as a
  truncated geometric series since the emissivities are sufficiently
  close to unity and the view factors $F_{ij}<1$ with most of them in
  fact much smaller then unity. Hence the inverse kernel is given by
  $\bm{K}^{-1} \equiv \left[ \bm{1} - ( \bm{1} - \bm{\epsilon})
    \bm{F}\right]^{-1}=\sum_{n=0}^{n_c} \bm{M}^n$
  with $\bm{M}=( \bm{1} - \bm{\epsilon}) \bm{F}$. We find that $n_c=6$
  is sufficiently accurate approximation for the parameters used
  below.

\item Finally, we compute the surface patch temperatures $T_j$ by an iterative solution of the equilibrium condition 
  $\bm{q}^\text{ext} = \bm{q}^\text{int}$ [see Eq.~(\ref{eq:5})] for given surface emissivity $\epsilon_j$, downward radiation $L$,
  interior temperatures $T^\text{int}_j$ and effective thermal conductivity $\kappa_j/d_j$. The iteration steps are as follows:

\begin{itemize}

\item[(i)] Choose initial patch temperatures $T_j^{(\nu=0)}$, e.g.,  the internal temperature. Convergence to the same result is obtained for a wide range of initial choices. 
  
\item[(ii)] Compute the external flux
  $\bm{q}^{\text{ext}\, (\nu=0)} = (\bm{F} - \bm{1}) \bm{K}^{-1}
  \bm{J}^{(\nu=0)}_0$
  with the $N+1$ dimensional initial vector
  $\bm{J}^{(\nu=0)}_0=[L,\sigma \epsilon_1 {T_1^{(\nu=0)}}^4, \ldots ,
  \sigma \epsilon_N {T_N^{(\nu=0)}}^4]$.

\item[(iii)] Compute the updated patch temperatures $T_j^{(\nu=1)}$
  from the equation $q_j^{\text{ext}\, (\nu=0)} = ( T_j^{(\nu=1)} -
  T^\text{int}_j ) \kappa_j/d_j$ for $j=1,\ldots, N$.

\item[(iv)] Continue with step (i) to start the next iteration step,
  i.e.,
  $\bm{q}^{\text{ext}\, (\nu=1)} = (\bm{F} - \bm{1}) \bm{K}^{-1}
  \bm{J}^{(\nu=1)}_0$
  with the vector
  $\bm{J}^{(\nu=1)}_0=\{L,\sigma \epsilon_1 [(T_1^{(\nu=1)}+T_1^{(\nu=1)})/2]^4, \ldots ,
  \sigma \epsilon_N [(T_N^{(\nu=1)}+T_N^{(\nu=1)})/2]^4\}$.
\end{itemize}

In (iv) and all following iteration steps it is useful to use the average of the last two iterations for the patch temperatures, as indicated here, to obtain rapid convergence. Typically, for the models and parameters used below, after about 20 iterations a stable solution for the patch temperatures had been reached (within a relative accuracy of $10^{-4}$). 

\end{enumerate}

\bibliography{UHI}


\section*{Data Availability}

The datasets generated during and/or analysed during the current study are available from the corresponding author on reasonable request.

\section*{Acknowledgements}

Contributions of Jorge Gonzales, Prathap Ramamurthy, Brian Vant-Haul, David Sailor, Andreas Karpf, Gregory Dobler, Steve Koonin, David Messinger, and Julie Pullen are gratefully acknowledged.

\section*{Author Contributions}

M.G. designed the experimental campaign, M.A. and M.G.conducted the measurements and carried out the post processing, T.E. carried out model based computations and analyzed the results.  All authors wrote and reviewed the manuscript.

\section*{Additional Information}

\textbf{Competing Interests:}  The authors declare that they have no competing interests.

\end{document}